\documentclass[useAMS,usenatbib]{mn2e}
\usepackage{color,graphicx} 
\usepackage{latexsym, times, setspace, amssymb, amsfonts, rotating} 
\usepackage{natbib}
\usepackage[fleqn]{amsmath} 
\usepackage[integrals]{wasysym}
\usepackage[tight]{subfigure}

\usepackage{siunitx}  
\usepackage{lscape}

\begin{document}

\title[Long-term Observations of Three Nulling Pulsars]{Long-term Observations of Three Nulling Pulsars}

\date{\today}

\author[N.~J.~Young et al.]{N.~J.~Young,$^{1,2}$\thanks{E-mail:
    Neil.Young@wits.ac.za} P.~Weltevrede,$^3$ B.~W.~Stappers,$^3$
  A.~G.~Lyne$^3$ and M.~Kramer$^{3,4}$\vspace{0.4cm}\\ 
  \parbox{\textwidth}{$^1$SKA South Africa,
    4th Floor, The Park, Park Road, Pinelands 7405, South Africa\\ $^2$School
    of Physics, University of the Witwatersrand, PO BOX Wits, Johannesburg,
    2050, South Africa \\$^3$Jodrell Bank Centre for Astrophysics, School of
    Physics, The University of Manchester, Manchester M13 9PL, UK
    \\ $^4$Max-Planck-Institut f\"{u}r Radioastronomie, Auf dem H\"{u}gel 69,
    53121 Bonn, Germany}} \maketitle
\begin{abstract}
We present an analysis of approximately 200 hours of observations of the
pulsars J1634$-$5107, J1717$-$4054 and J1853$+$0505, taken over the course of
14.7~yr. We show that all of these objects exhibit long term nulls and
radio-emitting phases (i.e. minutes to many hours), as well as considerable
nulling fractions (NFs) in the range $\sim67\,\%-90\,\%$. PSR~J1717$-$4054 is
also found to exhibit short timescale nulls ($1-40~P$) and burst phases
($\lesssim200~P$) during its radio-emitting phases. This behaviour acts to
modulate the NF, and therefore the detection rate of the source, over
timescales of minutes. Furthermore, PSR~J1853$+$0505 is shown to exhibit a
weak emission state, in addition to its strong and null states, after
sufficient pulse integration. This further indicates that nulls may often only
represent transitions to weaker emission states which are below the
sensitivity thresholds of particular observing systems. In addition, we
detected a peak-to-peak variation of $33\pm1\,\%$ in the spin-down rate of
PSR~J1717$-$4054, over timescales of hundreds of days. However, no long-term
correlation with emission variation was found.

\end{abstract}

\section{Introduction}\label{sec:intro}
Pulsars are classically considered to be rapidly rotating neutron stars, which
radiate beamed radio emission from their poles in a steady and predictable
fashion. In this model, an observer on Earth receives a pulse every rotation
from a given pulsar, as its beam of emission sweeps across the observer's
line-of-sight (LOS). In reality, however, pulsars have been shown to exhibit
variability in their emission over every timescale which they can be observed
(i.e. from nanosecond bursts to multi-decadal variations; see, e.g.,
\citealt{hkwe03,kea13,lgw+13}).

Of particular interest is the phenomenon of pulse nulling, where the pulsed
emission from a pulsar appears to completely cease \citep{bac70a}, over
timescales of pulse periods ($P$) to many years (see, e.g.,
\citealt{wmj07,llm+12}). This can be caused by a number of scenarios: 1) a
pulsar could undergo complete cessation (a.k.a. deep nulls;
e.g. \citealt{klo+06,gjk12}) or 2) transition to a weaker emission mode, which
is below a given sensitivity threshold (e.g. \citealt{elg+05,yws+14}); 3) the
radio beam could also move out of the LOS due to unfavourable emission
geometry (e.g. \citealt{dzg05,mg06,zgd07,tim10}) or 4) the acceleration zone
might not be completely filled by electron-positron pairs, resulting in
time-dependent variations in an emission `carousel model' (see, e.g.,
\citealt{dr01,jv04,rw07}).

While the latter geometrical and beam-filling theories can account for short
nulls (i.e. $\lesssim10~P$), they cannot account for the typical
($\sim10^{1-2}~P$) to long-term ($\sim10^{3-8}~P$) nulls observed in many
objects (e.g. \citealt{klo+06,mll+06,crc+12}). In this context, nulls are
considered to be extreme manifestations of mode changes in pulsar emission,
due to global magnetospheric state changes \citep{wmj07,lhk+10}. This is
strongly supported by the observed correlations between emission variability
and spin-down ($\dot\nu$) rate changes in several sources
(e.g. \citealt{lhk+10,lst12a}), as well as the simultaneous X-ray$-$radio mode
switches observed in PSR~B0943$+$10 \citep{hhk+13}.  However, the triggers
responsible for such magnetospheric reconfigurations have still not been
firmly ascertained (e.g. \citealt{cs08,lhk+10,jon11,rmt11}).

Significant emphasis has been placed on discovering and monitoring pulsars
with extreme nulling and/or moding properties, which may give valuable insight
to determine the root cause of pulsar emission variability. Such objects may
undergo mode changes and/or nulls with long timescales
(e.g. \citealt{klo+06,ksj13,kek+13,bkb+14}). They may also exhibit large
nulling fractions (NFs; e.g. \citealt{bbj+11,gjk12}). However, very few
sources are known to display such features (see \citealt{bbj+11} for a
detailed discussion), which has somewhat stifled the amount of studies on
these enigmatic objects. Below, we shift our focus to three pulsars which have
been shown to exhibit extreme nulling behaviour.

The first pulsar we discuss, J1634$-$5107, is a 507~ms source which was
discovered in the Parkes Multibeam Pulsar Survey (PMPS;
\citealt{lfl+06}). Nulling activity in the source was first identified in
further analysis of a number of PMPS discoveries, where it was shown that the
pulsar exhibits a strong radio-emitting (i.e. `ON') state and a pure null
(i.e. `OFF') state with a periodicity of approximately 10~days
\citep{okl+06}. This analysis was later elaborated on by \cite{obr10}, who
analysed many hours of observations~$-$~recorded over hundreds of typically
short ($\lesssim30$~min) sessions~$-$~to infer a $\mathrm{NF}\sim86\,\%$ for
the source. Further investigation of the emission phases also showed that the
object does not fluctuate between null and emitting states on short timescales
(i.e. $t_{\mathrm{on}}\lesssim1$~d and $t_{\mathrm{off}}\lesssim9$~d).

PSR~J1717$-$4054, the second pulsar, is a 888~ms source which was discovered
in a Southern Galactic plane survey with the Parkes telescope
\citep{jlm+92}. Follow-up timing studies by \cite{jml+95} showed that the
source undergoes pulse nulling for $\sim75\,\%$ of the time. Consequently, the
number of detections were too few to enable a $\dot\nu$ measurement. This
source was also another subject of the \cite{okl+06} PMPS study, where a
slightly higher $\mathrm{NF}\sim80\,\%$ was inferred. In this study, it was
noted that the source transitions to and from null phases over timescales less
than a second, indicating an abrupt cessation mechanism. Following this work,
\cite{wmj07} presented the analysis of a single 2-h observation of the
source. They detected the object for the first 210~s of the observation, after
which the object transitioned to, and remained in, its null state. As a
result, they inferred a $\mathrm{NF}\gtrsim95\,\%$ and transition
timescale~$-$~the expected length of time between consecutive mode
transitions~$-$~of $\gtrsim2$~h. This result was later refined by
\cite{obr10}, who inferred an average $\mathrm{NF}\sim74\,\%$ from the
analysis of 252 observations. They also reported the presence of short
timescale nulls during active emission phases.

The last pulsar, J1853$+$0505, is a relatively undocumented 905~ms source. It
was discovered in the PMPS \citep{hfs+04} and, since, has only been documented
for its pulse broadening characteristics due to interstellar scattering
\citep{bcc+04}. The nulling activity in this source was only identified
through further analysis of the PMPS discoveries and has remained unpublished
until now.

Since their initial discovery, PSRs~J1634$-$5107, J1717$-$4054 and
J1853$+$0505 have been routinely observed under several dedicated observing
programmes, predominantly with the Parkes telescope. This has facilitated more
comprehensive analysis of their emission variability and timing properties,
which is presented in this work. We note, however, that \cite{khs+14}
published results from an independent analysis of the same data set for
PSR~J1717$-$4054, during the preparation of this manuscript.

In the following section, we review the observing programmes used to gather
our data. We then discuss the emission, timing and polarimetric properties of
the objects, where possible, in sections \ref{sec:em_prop}, \ref{sec:timing}
and \ref{sec:pol}, respectively. Lastly, we discuss the implications of our
results in section~\ref{sec:conc} and compare them with the results of
\cite{khs+14} for PSR~J1717$-$4054.

\section{Observations}\label{sec:obs}
Observations of PSRs~J1634$-$5107 and J1717$-$4054 were obtained through an
intermittent source monitoring programme (IMP) which was conducted with the
Parkes~\hbox{64-m} radio telescope, over the period of 1999~August~21 to
2014~May~12. The majority of these data were obtained with the H-OH receiver
and central beam of the Multibeam receiver. However, a number of observations
were also obtained using the 10/50cm receiver (refer to Table~\ref{tab:obs};
see also \citealt{mhb+13} for detailed specifications on the Parkes
receivers).

An analogue filterbank (PAFB) system was primarily used to record the
observations for these pulsars up until 2008~September, after which digital
filterbank (PDFB) systems were used to obtain the majority of the remaining
data (see Table~\ref{tab:obs}). Note that regular observations of the pulsars
with the PDFBs commenced from 2007~December. A total of 14 observations were
also recorded with the ATNF Parkes Swinburne Recorder, CASPER Parkes Swinburne
Recorder and Wide Band Analogue Correlator backends for these
pulsars\footnote{Refer to
  http://astronomy.swinburne.edu.au/pulsar/?topic=instrumentation and
  \cite{mhb+13} for more details on Parkes backends.}.

\begin{table*}
\caption{The observation characteristics of the observing programmes conducted
  for each source. PSR denotes the pulsar observed, REC denotes the receiver/
  backend reference, MJD refers to the modified Julian Date at the start of
  the observations and $T_{\mathrm{span}}$ represents the total time-span of
  the observations. The total number of observations carried out is given by
  $N_{\mathrm{obs}}$, $T_{\mathrm{obs}}$ denotes the typical observation
  duration and $\langle C\rangle$ represents the average observation
  cadence. The centre sky frequency, observation bandwidth and total number of
  frequency channels are also denoted by $\nu$, $\Delta \nu$ and
  $N_{\mathrm{chan}}$, respectively.}  \centering
\begin{tabular}{ l l l l l l l l l l }
  \hline
   PSR & REC & MJD & $T_{\mathrm{span}}$ (d) & $N_{\mathrm{obs}}$ &
   $T_{\mathrm{obs}}$ (min) & $\langle C\rangle$ (week$^{-1}$) &  $\nu$ (MHz)
   & $\Delta \nu$ (MHz) & $N_{\mathrm{chan}}$\\
  \hline
  J1634$-$5107 & Multi   & 51411.4  & 5378.2 & 383 & 7.6   & 0.5  & 1374 & 288 & 96 \\
               & H-OH    & 52961.1  & 281.2  & 73  & 10.3  & 1.8  & 1518 & 576 & 96 \\
               & 10/50cm & 53000.2  & 609.0  & 35  & 10.7  & 0.4  & 2934/685 & 576/64  & 192/256\\\hline

  J1717$-$4054 & Multi   & 53280.3  & 3509 & 335 & 5.5   & 0.7  & 1374 & 288  & 96     \\
               & H-OH    & 54112.0  & 116  & 21  & 14.3  & 1.3  & 1518 & 576  & 1024   \\
               & 10/50cm & 53029.8  & 3181 & 15  & 5.0   & 0.03 & 3094/732 & 1024/64 & 1024 \\\hline
 
  J1853$+$0505 & Multi   & 51633.8  & 3212.8 & 94  & 8.6  & 0.2  & 1374 & 288  & 96   \\
               & H-OH    & 53150.6  & 90.9   & 19  & 14.5 & 1.5  & 1518 & 576  & 1024 \\
               & 10cm    & 53190.5  & 1.2    & 6   & 15.0 & 35.0 & 685  & 64   & 1024 \\
               & LAFB    & 51843.4  & 3527.7 & 250 & 16.5 & 0.5  & 1396 & 32   & 32   \\
               & LDFB    & 54846.6  & 829.7  & 23  & 11.8 & 0.2  & 1520 & 384  & 768  \\
  \hline
\end{tabular}
\label{tab:obs}
\end{table*}

The PAFB observations were one-bit digitised at 250~$\mu$s intervals and were
later folded offline to produce both folded data, with typically 256 bins per
period at 1~min sub-integration intervals, as well as single-pulse archives
where possible\footnote{Single-pulse data was obtained for 331 observations of
  PSR~J1634$-$5107, 205 observations of PSR~J1717$-$4054 and 90 observations
  of PSR~J1853$+$0505, respectively.}. The PDFB observations were folded
online to typically provide 512 bins per period, using 8-bit digitisation, at
1~min sub-integration intervals.  A polarised calibration signal was also
injected into the receiver probes, and observed, prior to a large number of
these observations with the Multibeam receiver, in order to polarisation
calibrate a subset of the data.

Observations of PSR~J1853$+$0505 were obtained with both the
Parkes~\hbox{64-m} and Lovell~\hbox{76-m} radio telescopes, over the period of
2000~March~30 to 2014~May~12 under a joint IMP. The same receivers and
observing setup as described above was used to obtain observations of this
source at Parkes. The Lovell observations were obtained with a 20-cm
dual-orthogonal, linear feed receiver, coupled with an analogue filterbank
(LAFB) or digital filterbank (LDFB). The LAFB data were folded online to
provide 400 bins per period at 1~min sub-integration intervals, using one-bit
digitisation. The LDFB observations were folded online to provide 1024 bins
per period at 1~min sub-integration intervals, using 8-bit digitisation (refer
to Table~\ref{tab:obs}).

In off-line processing, we de-dispersed and examined the data for
radio-frequency interference (RFI). For the Parkes and LDFB data, we carried
out manual RFI mitigation through use of {\small\textsc{PSRZAP}} and
{\small\textsc{PAZ}}\footnote{See \cite{hvm04} and
  http://psrchive.sourceforge.net/manuals for details on the
  {\small\textsc{PSRCHIVE}} software suite.} so as to reduce the number of
frequency channels and single pulses/sub-integrations excised from the
data. We also reduced the number of bins to 256 per period for each of these
observations, for the emission variability analysis. Whereas, we reduced the
number of bins to 150 per period for the polarimetric analysis.

For the LAFB data, we were restricted to only 32~MHz of bandwidth. Therefore,
limited manual RFI excision was only possible for these data. The LAFB data
were also converted into the
{\small\textsc{SIGPROC}}\footnote{http://sigproc.sourceforge.net/} file
format, to facilitate variability analysis of the source.

\section{Emission Properties}\label{sec:em_prop}
\subsection{Variability and Emission Timescales}\label{sec:variability}
Here we describe the nulling activity of the three pulsars chosen for our
study. For each of these objects, the inferred scintillation bandwidth from
the {\small\textsc{NE2001}} model is less than 2~kHz \citep{cl02}. This is
consistent with our analysis of their dynamic spectra, which do not exhibit
any prominent scintles given the frequency resolution afforded
(i.e. $\gtrsim500$~kHz). Thus, the variability of these objects cannot be
described by propagation effects in the interstellar medium
(e.g. scintillation). Rather, the nulling behaviour described below is
considered to be intrinsic to the sources.

\subsubsection{PSR~J1634$-$5107}\label{sec:j1634var}
PSR~J1634$-$5107 was detected in 64 out of the 491 observations analysed,
which amounts to a total of $\sim$7.0~h in the ON state from the $\sim$67.7~h
data set. No weak or underlying emission was observed during individual OFF
observations, or in the average OFF profile for each observing frequency
(c.f. B0826$-$34; \citealt{elg+05}), which is consistent with previous
findings. Moreover, we discovered that the detectability of the source is
strongly correlated with integration length. That is, we found that
sub-integration lengths of $\lesssim2$~min do not provide sufficient
sensitivity for reliable characterisation of mode transitions due to apparent
nulls. This can be explained by the low flux density of the source
(c.f. PSR~B1931$+$24; e.g. \citealt{klo+06,ysl+13}) and/or short timescale
nulling not probed by our data.

Therefore, we only comment on the properties of the object on an observation
to observation basis, where the mean length is $\sim$8~min. We subsequently
estimate a $\mathrm{NF}=90\pm5\,\%$ for the source~$-$~from the fractional
time it was observed in the OFF phase\footnote{Under the assumption that pulse
  nulling is a stochastic process, a NF can be accurately characterised from
  the fractional time observed in an OFF phase over a sufficient number of
  observations.}~$-$~which is consistent with the result of \cite{obr10}. Note
that we quote the uncertainty on the NF assuming Poissonian sampling
statistics. That is, we assume that each measurement of the source in a
particular emission state is independent. Thus, we estimate the fractional
error on the NF as $\sim\sqrt{N_{\mathrm{obs}}}/N_{\mathrm{obs}}$. We stress,
however, that this theoretical estimate will not be as accurate as a
constraint obtained from documenting a large number of emission phases
continuously.

\begin{figure}
  \centering
    \includegraphics[trim= 11mm 0mm 3mm 12mm,clip,angle=270,totalheight=15in,height=5cm,width=8.25cm]{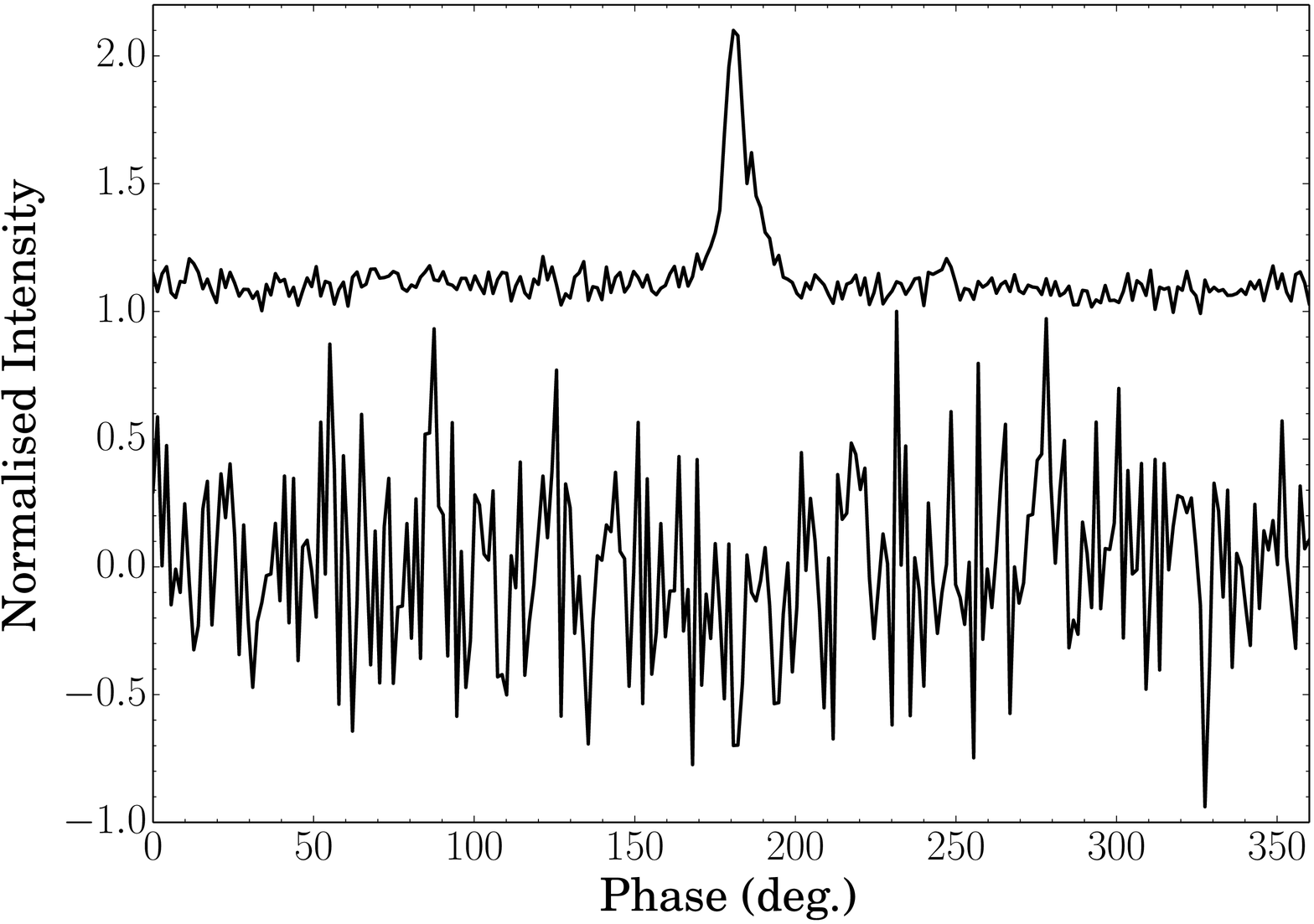}
\vspace{-1mm}
\caption{Average emission profiles for PSR~J1634$-$5107 from the longest
  observations, separated by emission mode and normalised by their peak
  intensities. \emph{Bottom:} OFF profile from an 4.5~h observation taken on
  2002~September~13 at 04:52:05 UTC. \emph{Top:} ON profile from a 33~min
  observation recorded on 2002~January~19 at 19:12:26 UTC, offset vertically
  for clarity. Note that the flux density of the OFF emission is
  $\lesssim1.7\pm0.5\,\%$ that of the ON emission for these data.}
 \label{1634avprofs}
\end{figure}

Further to the above, we did not observe any mode transitions during
individual observations of the source. As such, the maximum observation
durations of 4~h~30~min for the OFF mode and 33~min for the ON mode (see
Fig.~\ref{1634avprofs}) do not accurately constrain the source's emission
timescales. Instead, we place constraints on the emission timescales through
analysis of the longest and highest cadence observing runs.

In our data set, there were two particularly long (i.e. 10~h) observing runs
(between 2002~November~1 and 2002~November~3) where the pulsar was undetected
throughout each half-hour to hourly observation. By comparison, there was one
particularly long (i.e. 6.5~h) observing run on 2004~May~25 where the source
was detected throughout each half-hour to hourly observation.  Further to
this, we place tentative limits on the maximum emission timescales from a
series of observations, between 2003~August~12 and 2003~August~16, where the
pulsar was seen to switch between modes on a daily basis. This indicates that
the source undergoes relatively long emission phases, which are at least
several hours up to a day in length.

\subsubsection{PSR~J1717$-$4054}\label{sec:j1717var}
PSR~J1717$-$4054 was detected in 119 out of the 371 observations analysed, and
was observed to transition between its ON and OFF modes on 38
occasions. Analysis of the sub-integration data for each observation showed
that the source was ON for a total of 8.6~h of the 36.7~h data set. This
equates to a $\mathrm{NF}=77\pm5\,\%$~$-$~using the error estimation method
presented in $\S$~\ref{sec:j1634var}~$-$~which is consistent with the findings
of both \cite{jml+95} and \cite{okl+06}. We also infer the average transition
timescale to be $\sim$56~min from the total observation length and number of
fully resolved mode transitions ($T_{\mathrm{obs}}/N_{\mathrm{switch}}$).

\begin{figure}
  \centering
    \includegraphics[trim= 0mm 11mm 20mm 30mm,clip,angle=0,totalheight=10cm,height=6cm,width=8.4cm]{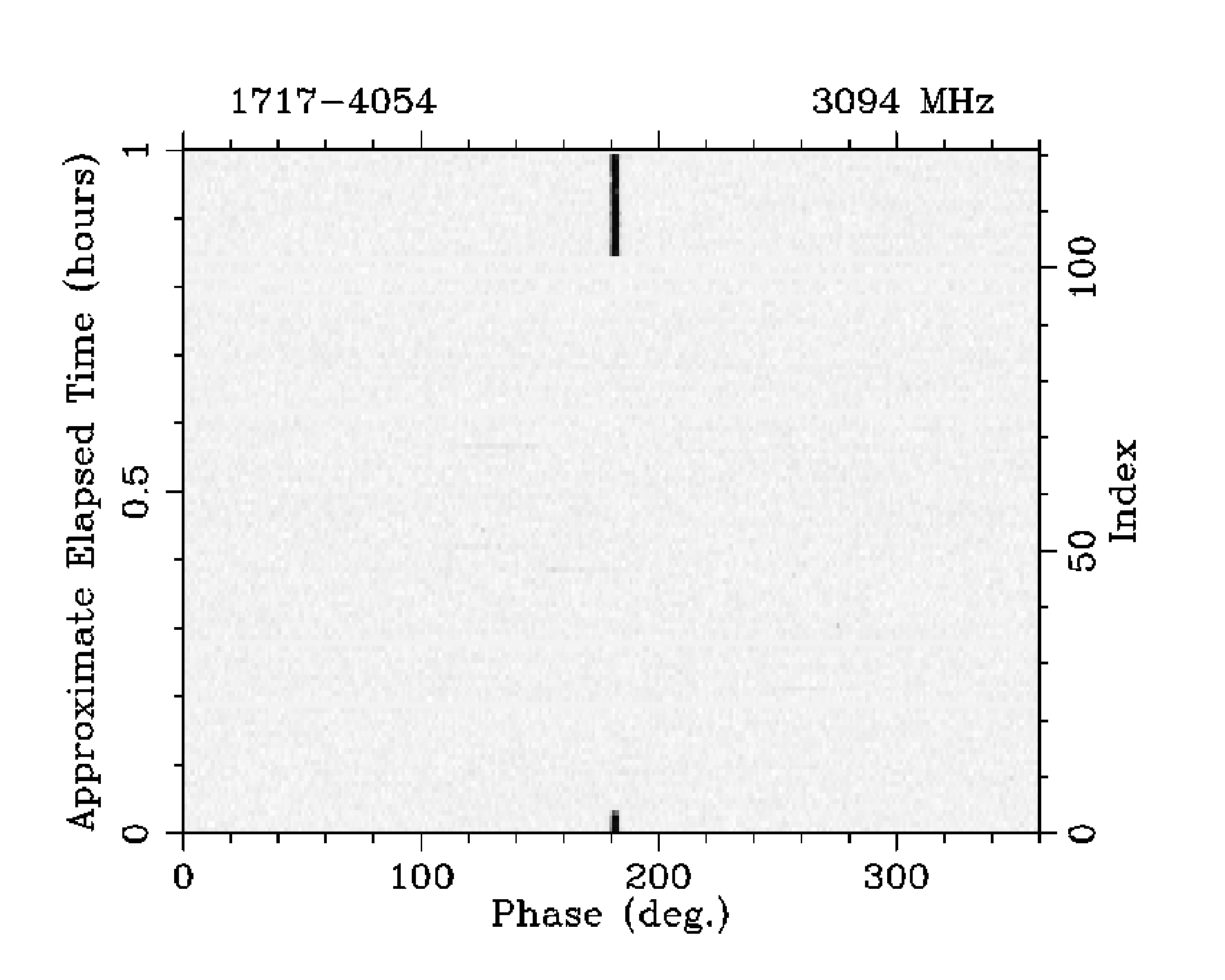}
    \includegraphics[trim= 11mm 0mm 3mm 12mm,clip,angle=270,totalheight=15in,height=5cm,width=8.25cm]{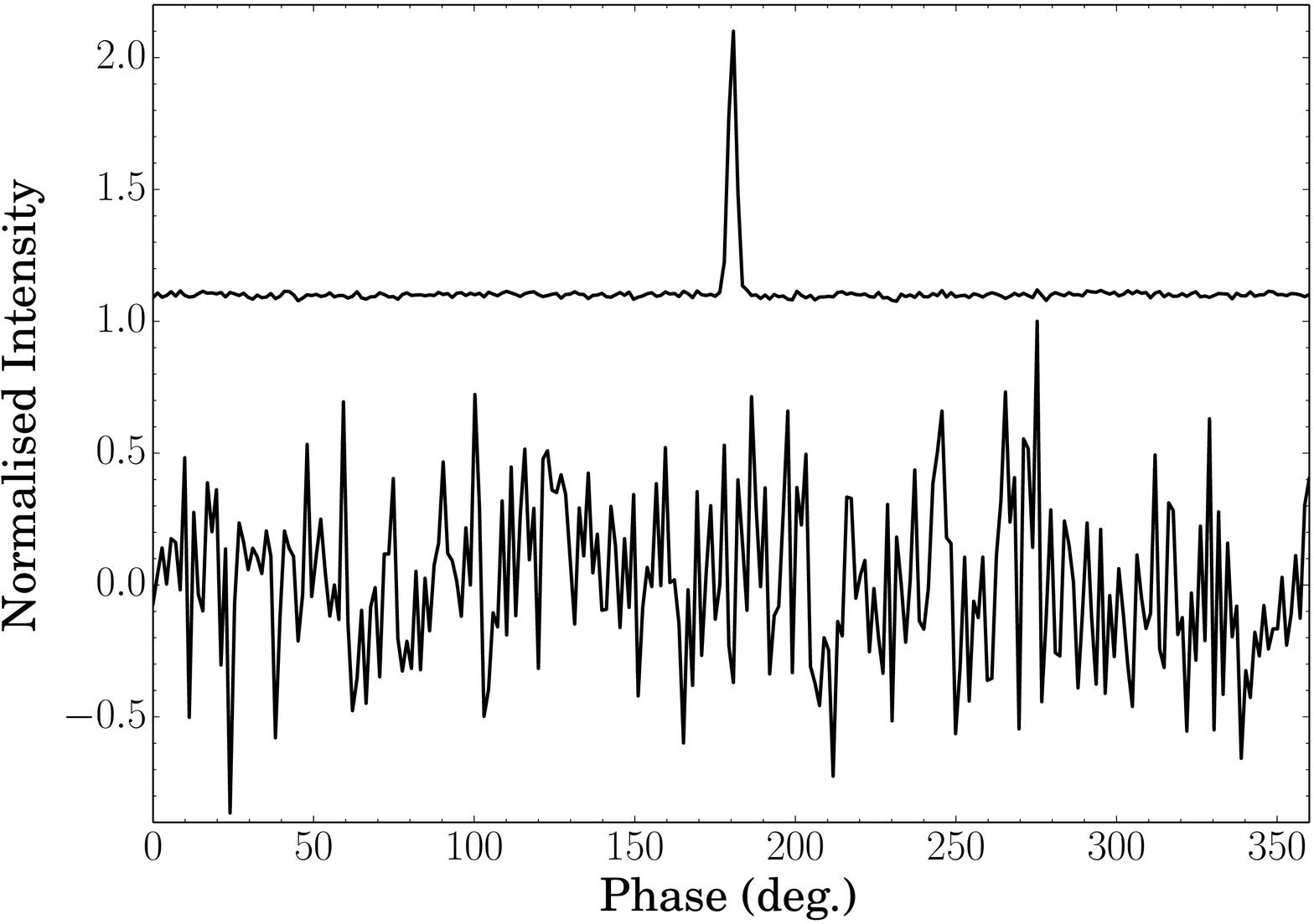}
\vspace{-1mm}
\caption{An hour-long observation of PSR~J1717$-$4054 recorded on
  2012~October~10 at a centre frequency of 3094~MHz. \emph{Top panel:}
  Sub-integration versus phase intensity map showing the pulsar transition
  between the ON and OFF modes and back again. \emph{Bottom panel:} Average
  profiles for the corresponding OFF (bottom) and ON (top) sub-integrations
  displayed in the top panel, normalised by their peak intensities and offset
  vertically for clarity. Note that no emission is detected through
  integration of 48~min of data. The null confusion limit is also
  $\lesssim0.8\pm0.2\,\%$ of the flux density of the ON emission for these
  data.}
 \label{1717transitions}
\end{figure}

The null lengths in the sub-integration data were observed to span from
0.5~to~$\gtrsim64$~min. Whereas, the ON phases were observed to last between
0.5 and 12.5~min in individual observations. Given the poor constraints on the
upper limits of each emission timescale, we also analysed successive
contiguous observations which were not separated by more than 10~min. From
this analysis, we infer a maximum ON timescale of 16.2~min and the same
maximum OFF timescale of $\gtrsim64$~min. Note that this analysis does not
completely constrain the possible range of ON timescales due to inadequate
observation cadence. Therefore, it is possible the maximum ON timescale could
be longer than that inferred from this work. The variability of the source is
demonstrated in an hour-long observation shown in Fig.~\ref{1717transitions}.

Upon inspection of the available single-pulse data, we were also able to
confirm that the source transitions between emission modes over timescales of
$\lesssim1\,P$. In addition, we found that the source exhibits a bimodal
distribution of nulls. That is, the source preferentially undergoes nulls over
timescales of $\sim1-40\,P$ or $\gtrsim340\,P$. We also note that emission
bursts from the pulsar only last $\sim1-200\,P$. Therefore, it is clear that
the majority of the short timescale nulls will not be resolved in the
sub-integration data of the ON phases where
$t_{\mathrm{sub}}\gtrsim60$~s. Thus, the interpretation of the ON emission
timescales is dependent on the observation setup.

The above results are consistent with the results of \cite{obr10}, but are
markedly different from the NF estimate of \cite{wmj07} taken from a single
observation. We attribute this to variation in the OFF and ON timescales,
where the wide range of activity timescales observed in our data set can lead
to variation in NF estimates between observations.

\subsubsection{PSR~J1853$+$0505}\label{sec:1853var}
\begin{figure}
  \centering
    \includegraphics[trim= 0mm 11mm 20mm 30mm,clip,angle=0,totalheight=10cm,height=6cm,width=8.4cm]{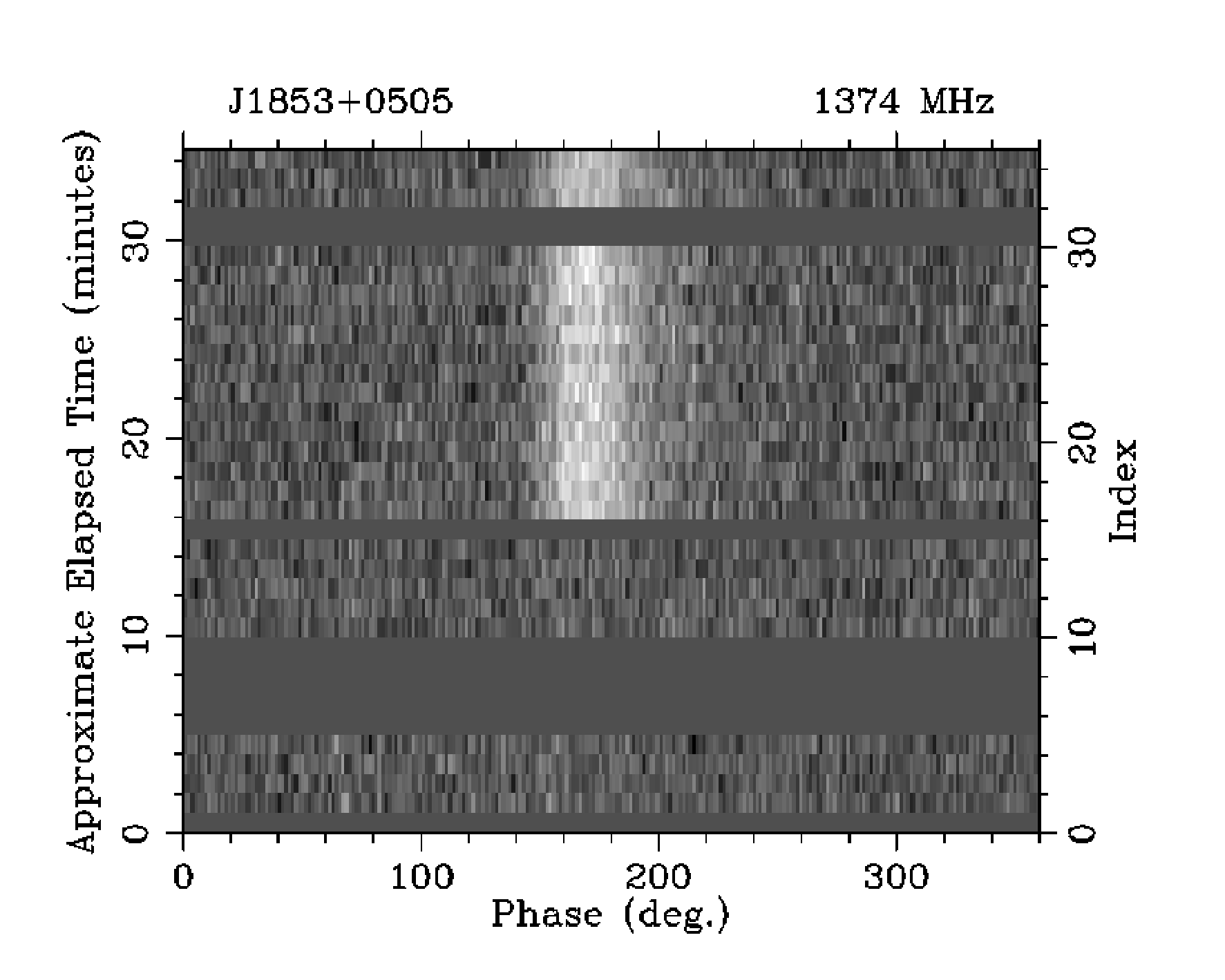}
    \includegraphics[trim= 11mm 0mm 3mm 12mm,clip,angle=270,totalheight=15in,height=5cm,width=8.25cm]{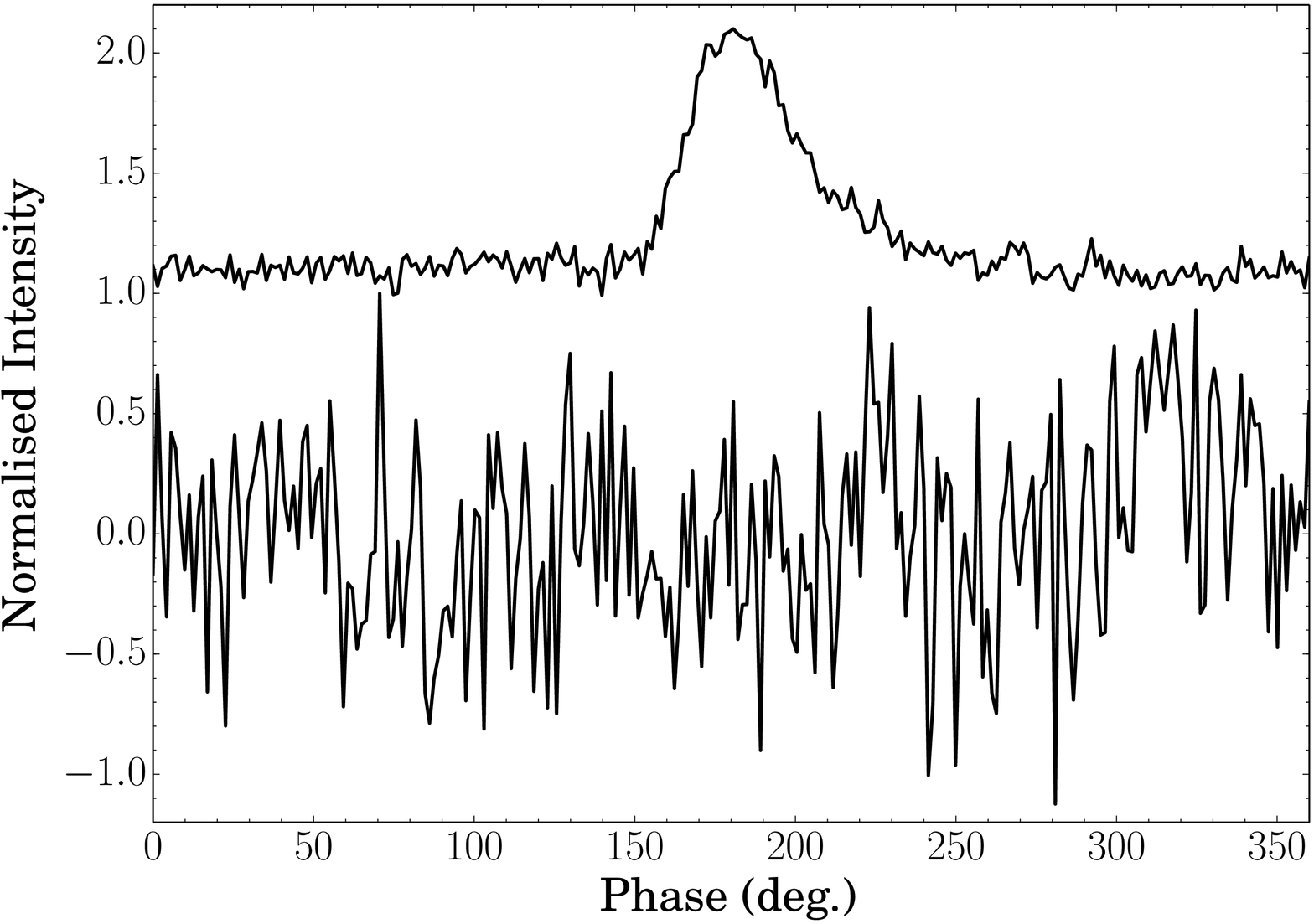}
\vspace{-1mm}
\caption{A 34.4~min long observation of PSR~J1853$+$0505 recorded on
  2001~June~17 at a centre frequency of 1374~MHz. \emph{Top panel:}
  Sub-integration versus phase intensity map showing the pulsar transition
  between the OFF and ON modes. Note that the blank horizontal lines represent
  sub-integrations weighted to zero from the RFI excision
  process. \emph{Bottom panel:} Average profiles for the corresponding OFF
  (bottom) and ON (top) sub-integrations displayed in the top panel,
  normalised by their peak intensities and offset vertically for clarity. Note
  that no emission is detected through integration of 9~min of OFF data. The
  null confusion limit is also $\lesssim3\pm1\,\%$ of the ON emission for
  these data.}
 \label{1853transitions}
\end{figure}

PSR~J1853$+$0505 was detected in 93 out of the 392 observations analysed, and
was observed to transition between its ON and OFF states on 29 occasions in
individual observing runs (see, e.g., Fig.~\ref{1853transitions}). Analysis of
the sub-integration data showed that the pulsar was ON for a total of 17.8~h
of the 92.8~h combined data set. For the LAFB data set we further integrated
observations to sub-integration times of $\gtrsim6$~min, to counter the low
bandwidth and subsequent reduction in sensitivity. Whereas, we were still able
to probe typical timescales of $\gtrsim1$~min for the Parkes and LDFB data
given the greater sensitivity.

\begin{figure}
  \centering
    \includegraphics[trim= 2mm -1mm 2mm -4mm,clip,angle=0,totalheight=10cm,height=5.75cm,width=8cm]{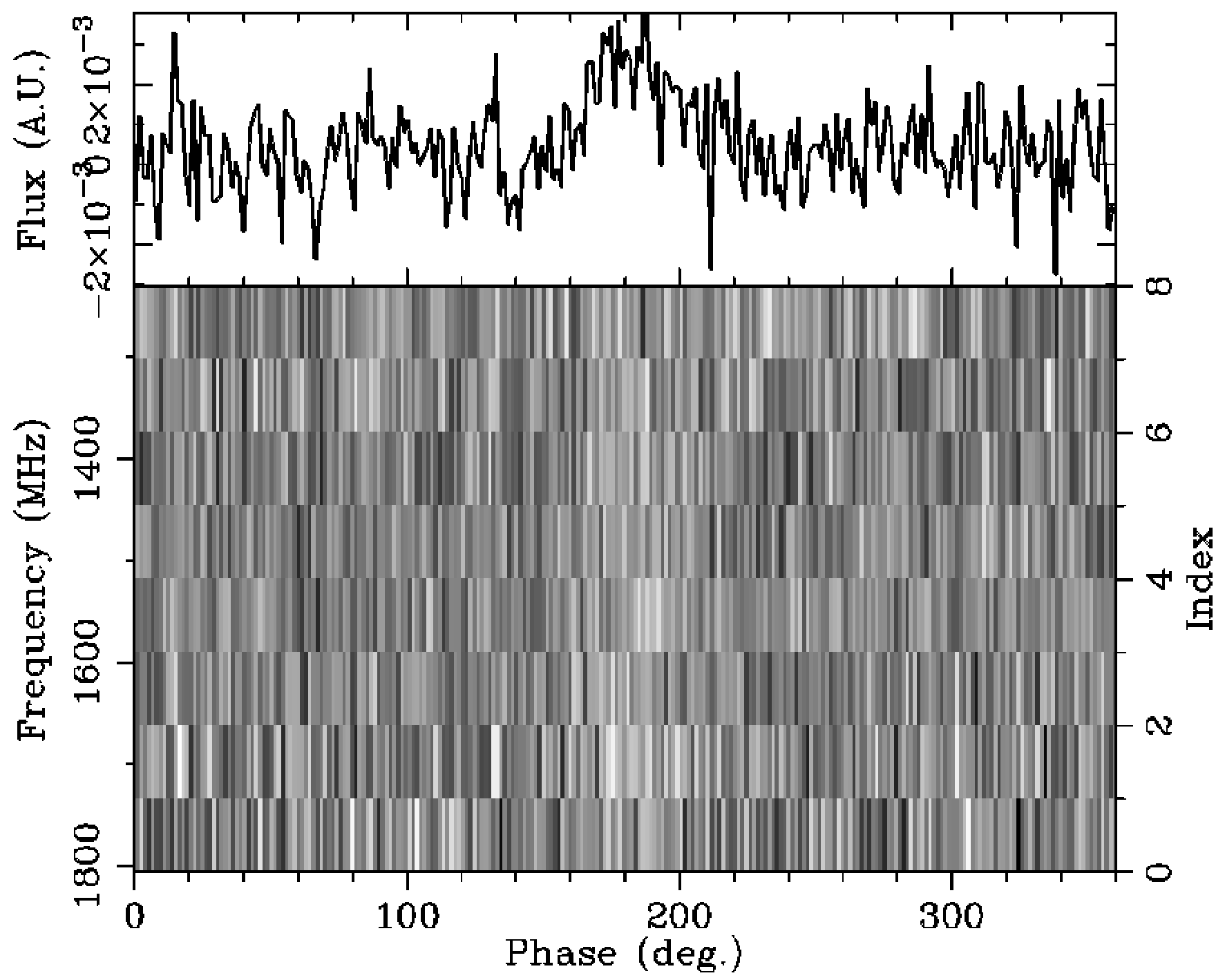}
    \includegraphics[trim= 2mm -4mm 2mm -1mm,clip,angle=0,totalheight=10cm,height=5.75cm,width=8cm]{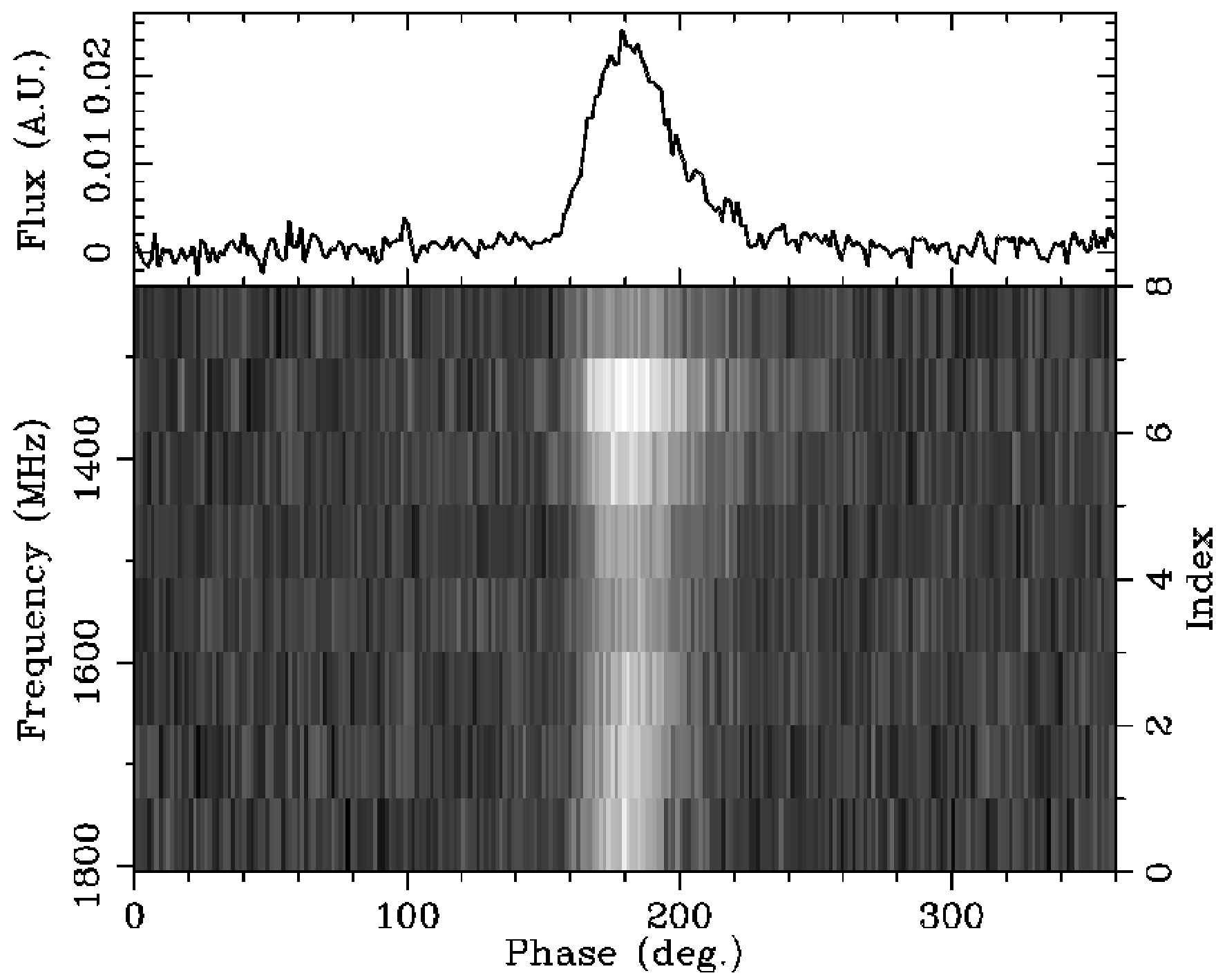}
\vspace{-3mm}
\caption{Two 15~min observations of PSR~J1853$+$0505, obtained at 1518~MHz,
  demonstrating the different emission states of the source. The \emph{top
    plot} shows a weak detection of the source, which was obtained an hour
  before an OFF detection and 2 hours before the strong detection, shown in
  the \emph{bottom plot}. The integrated profile and frequency versus phase
  intensity plot is displayed for each observation, in the top and bottom
  panels of each plot, respectively.}
 \label{UE}
\end{figure}

In addition to the strong detections, we noted that the source appears
extremely weak in 12 of the observations (refer to Fig.~\ref{UE}). This weak
emission is not detected over timescales of less than several minutes. Rather,
it can only be observed through integration of sufficiently long observations
(i.e. $\gtrsim10$~min), which is similar to the underlying emission observed
in a number of sources (see, e.g., \citealt{elg+05,wmj07,yws+14,syh+14}).

Remarkably, weak-mode transitions were also observed in two observations,
spaced over a year apart. In the first observation, the pulsar exhibited
weak-emission for $\sim16$~min, then abruptly transitioned to its strong mode
(i.e. between $\sim1$~min sub-integrations) for the remaining $\sim19$~min of
the observation. In the other observation, the source emitted in its strong
mode for $\lesssim1$~min, then abruptly switched to its weak mode for the
remaining $\sim9$~min. These attributes indicate that the weak emission mode
is relatively stable, which occurs between strong and OFF phases of
emission. The above also suggests a common driving mechanism for the separate
emission modes, due to the consistency of their transition timescales.

Although the LAFB data was subject to more RFI, compared with data obtained
with the other backends, the longer observing runs recorded provide greater
constraints on the cumulative ON and OFF timescales of the source. In these
data, we did not resolve any ON modes of less than 30~min in duration, nor did
we observe any greater than 2~h in length. Similarly, the minimum and maximum
OFF timescales were found to be 40~min and $\gtrsim7$~h~$20$~min,
respectively. However, we emphasise caution in assuming the upper limit for
the maximum OFF length, given that the LAFB data were limited in being able to
probe weak-mode detections.

We calculated NF estimates for both the LAFB ($\mathrm{NF}_{_{\mathrm{LAFB}}}$
$=$ $86\,\pm6\,\%$) and combined Parkes-LDFB ($\mathrm{NF}_{_{\mathrm{cmbd}}}$
$=$ $67\,\pm\,8\,\%$) data sets separately, due to the different observation
sensitivities. Considering the prevalence of weak emission in the combined
Parkes and LDFB data set, i.e. $\sim$11$\,\%$ of the total time, we suspect
that the NF estimate for the LAFB data is most likely overestimated. As such,
we assume the NF from the combined Parkes-LDFB data provides a better
representation of the variability of the source. We also infer an average
timescale of $\sim$3.2~h between contiguous ON modes from the total
observation length and number of observed transitions.

\subsection{Flux Density Limits}\label{sec:flux}
The detection of particularly weak emission in a number of objects~$-$~through
sufficient pulse integration (see, e.g., \citealt{elg+05,wmj07,yws+14}) or
migration to lower observing frequencies \citep{syh+14}~$-$~raises speculation
as to whether all nulls actually represent weak emission states or not. This
is punctuated by the fact that every telescope has a specific flux density
limit, for a given integration time, which may or may not allow the detection
of such low intensity emission states. With the above in mind, we sought to
place upper limits on the flux densities of the separate emission modes
($S_{\mathrm{off}}:S_{\mathrm{on}}$) for the pulsars in this study, so as to
characterise the possibility of null confusion in our data.

To estimate the flux density of the sources, during their separate emission
modes, we formed time- and frequency-averaged profiles for corresponding
observations obtained with the Parkes telescope. Observations were aligned
using the timing solution for each source, prior to integration. We then used
the modified radiometer equation to estimate the average flux densities
attributed to each mode from the mean signal-to-noise ratios (SNRs; see, e.g.,
\citealt{lk05}). Here, we assume that the dominant uncertainties arise from
gain and system temperature variations with respect to elevation for a given
observing session. We thus conservatively assume a $20\,\%$ variation in flux
density between each observation. We note, however, that the actual
uncertainites on our flux measurements are likely to be lower than this upper
limit, especially for the 10/50cm observations which were obtained in close
succession.  Table~\ref{tab:flux} shows the result of this analysis for the
three sources.

\begin{table*}
\caption{Average pulse properties for PSRs~J1634$-$5107, J1717$-$4054 and
  J1853$+$0505. The pulsar J2000 designation, observation frequency,
  bandwidth, number of observations used for each fold and total integration
  length are denoted by PSR~J, $\nu$, $\Delta \nu$, $N_{\mathrm{on,off}}$ and
  $T_{\mathrm{on,off}}$, respectively. The mean signal-to-noise ratio, average
  equivalent width and flux density are represented by $\mathrm{SNR}$,
  $W_{\mathrm{eq}}$ and $S_{\mathrm{on,off}}$, respectively. The upper limit
  on the OFF:ON flux density is also denoted by
  $S_{\mathrm{off}}:S_{\mathrm{on}}$. Note that we assume an upper limit of
  $\mathrm{SNR}=3$ for the OFF emission. The uncertainties in the flux
  densities are also quoted in parentheses with respect to the least
  significant digit.}  \centering
\begin{tabular}{ l l l l l l l l l l l l }
  \hline
   PSR & $\nu$ & $\Delta \nu$ & $N_{\mathrm{on}}$ & $N_{\mathrm{off}}$ &
   $T_{\mathrm{on}}$ & $T_{\mathrm{off}}$ & $\mathrm{SNR}$ & $W_{\mathrm{eq}}$
   & $S_{\mathrm{on}}$ & $S_{\mathrm{off}}$ & $S_{\mathrm{off}}:S_{\mathrm{on}}$\\
   & (MHz) & (MHz) & & & (s) & (s) & & (ms) & (mJy) & ($\mu$Jy) & ($10^{-3}$)\\
  \hline
  J1634$-$5107 & 1374 & 288  & 26 & 136 & 10986.8 & 85650.4 & 161.0  & 14.4 &
  0.40 (8) & $\lesssim2.6$~(5) & $\lesssim7$~(2) \\
              & 1518 & 576  & 13 & 27  & 7488.0  & 17072.6 & 119.4  & 15.7 &
  0.22 (4) & $\lesssim3.6$~(8) & $\lesssim17$~(5) \\\hline

  J1717$-$4054 & 732  & 64    & 2  & 2   & 874.9   & 3784.9  & 132.5  & 69.6 &
  7 (1) & $\lesssim80$~(20) & $\lesssim11$~(3) \\   
               & 1374 & 288  & 44 & 55  & 12152.8 & 16104.4  & 1393.8 & 13.3 &
  2.3 (5)  & $\lesssim4.4$~(9) & $\lesssim1.9$~(6) \\
               & 1518 & 576  & 8  & 7   & 2156.5  & 2096.6   & 406.5 & 11.6 &
  0.9 (2)   & $\lesssim7$~(1) & $\lesssim7$~(2) \\
               & 3094 & 1024  & 2  & 2   & 874.9   & 3963.3  & 158.2 & 7.8  &
  0.6 (1)   & $\lesssim5$~(1) & $\lesssim9$~(3) \\\hline

  J1853$+$0505 & 1374 & 288  & 27 & 11  & 16466.9 & 3768.0  & 305.8  & 105.7 &
  1.3 (3) & $\lesssim27$~(5) & $\lesssim20$~(6) \\
               & 1518 & 576  & 10  & 1   & 8985.6  & 898.6   & 190.9  & 93.2  &
  0.6 (1)  & $\lesssim30$~(6) & $\lesssim50$~(20) \\
  \hline
\end{tabular}
\label{tab:flux}
\end{table*}

Both PSRs~J1634$-$5107 and J1853$+$0505 were not observed during their ON
modes while the 10/50cm receiver was in use. Therefore, we cannot provide any
constraints on the ON or OFF average flux densities in either the 10- or 50-cm
band for these sources.

Comparing the flux densities of PSRs~J1634$-$5107, J1717$-$4054 and
J1853$+$0505, we place upper limits on their
$S_{\mathrm{off}}:S_{\mathrm{on}}$ values of $\lesssim(7\pm2)\times10^{-3}$,
$\lesssim(1.9\pm0.6)\times10^{-3}$ and $\lesssim(2\pm0.6)\times10^{-3}$,
respectively. The stringent null confusion limit placed on PSR~J1717$-$4054
indicates that the source could undergo deep nulls which are well below the
sensitivity threshold of the Parkes observing system. Whereas, the limits
placed on PSR~J1634$-$5107 and particularly PSR~J1853$+$0505 indicates that
null confusion could occur in these sources. For PSR~J1853$+$0505, this is
consistent with its observed weak emission, which is detected just above the
noise level (see also $\S$~\ref{sec:deepnull}).

\section{Timing}\label{sec:timing}
Considering the $\dot\nu$ variations observed in several moding and/or nulling
objects \citep{lhk+10}, we sought to characterise and compare the timing
properties of the pulsars in our study. Here, we computed the best-fit timing
solutions for the objects using the {\small\textsc{TEMPO2}}
package\footnote{An overview of this timing package is provided by
  \cite{hem06}.  See also http://www.atnf.csiro.au/research/pulsar/tempo2/ for
  more details.}  and the {\small\textsc{PSRCHIVE}} software
suite.\nocite{hmh+11}

\begin{figure}
  \centering
    \includegraphics[trim= 3mm 1mm -2mm 1mm,clip,angle=270,totalheight=10cm,height=4.5cm,width=8cm]{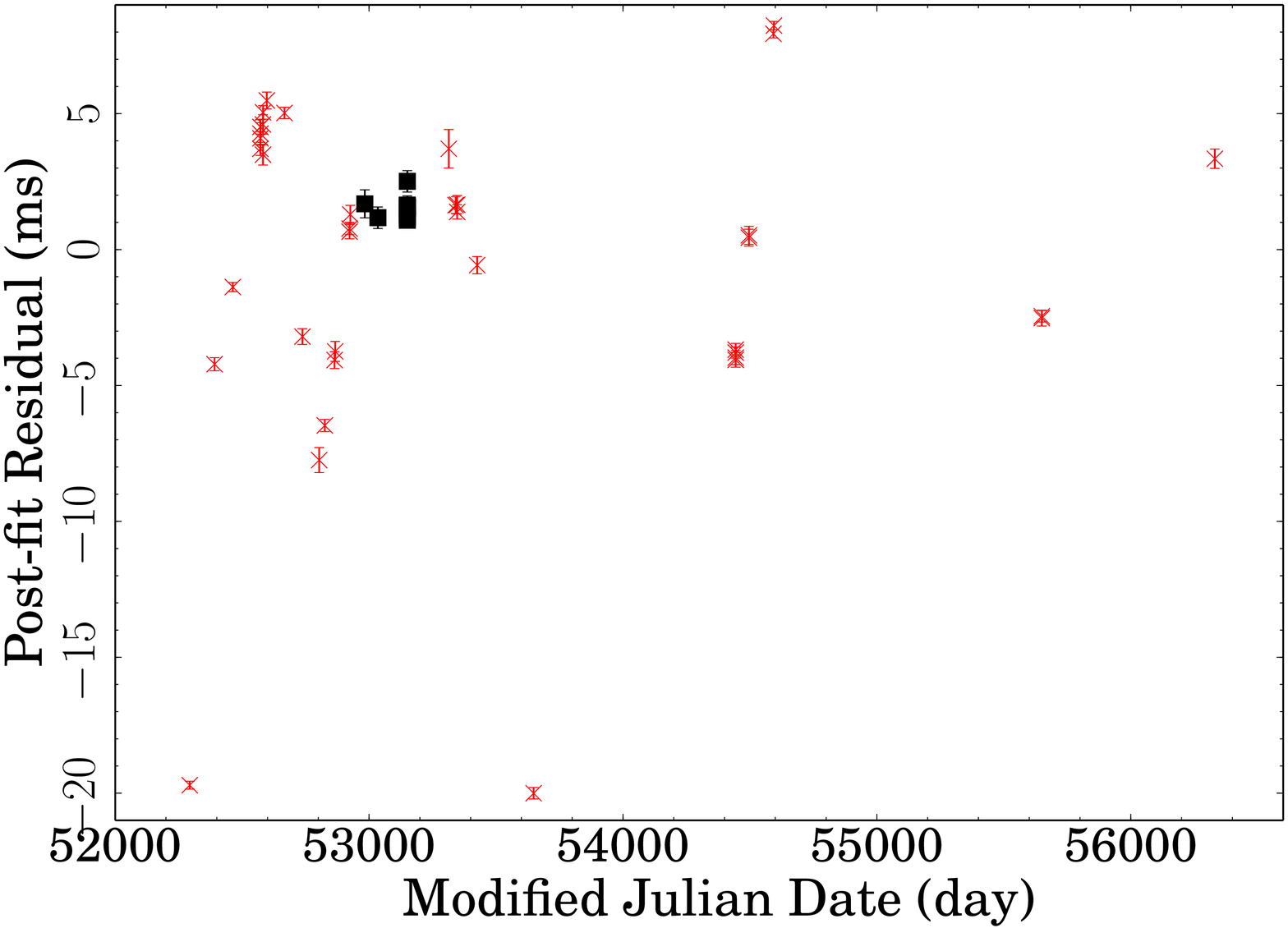}
    \includegraphics[trim= 3mm 1mm -2mm 1mm,clip,angle=270,totalheight=10cm,height=4.5cm,width=8cm]{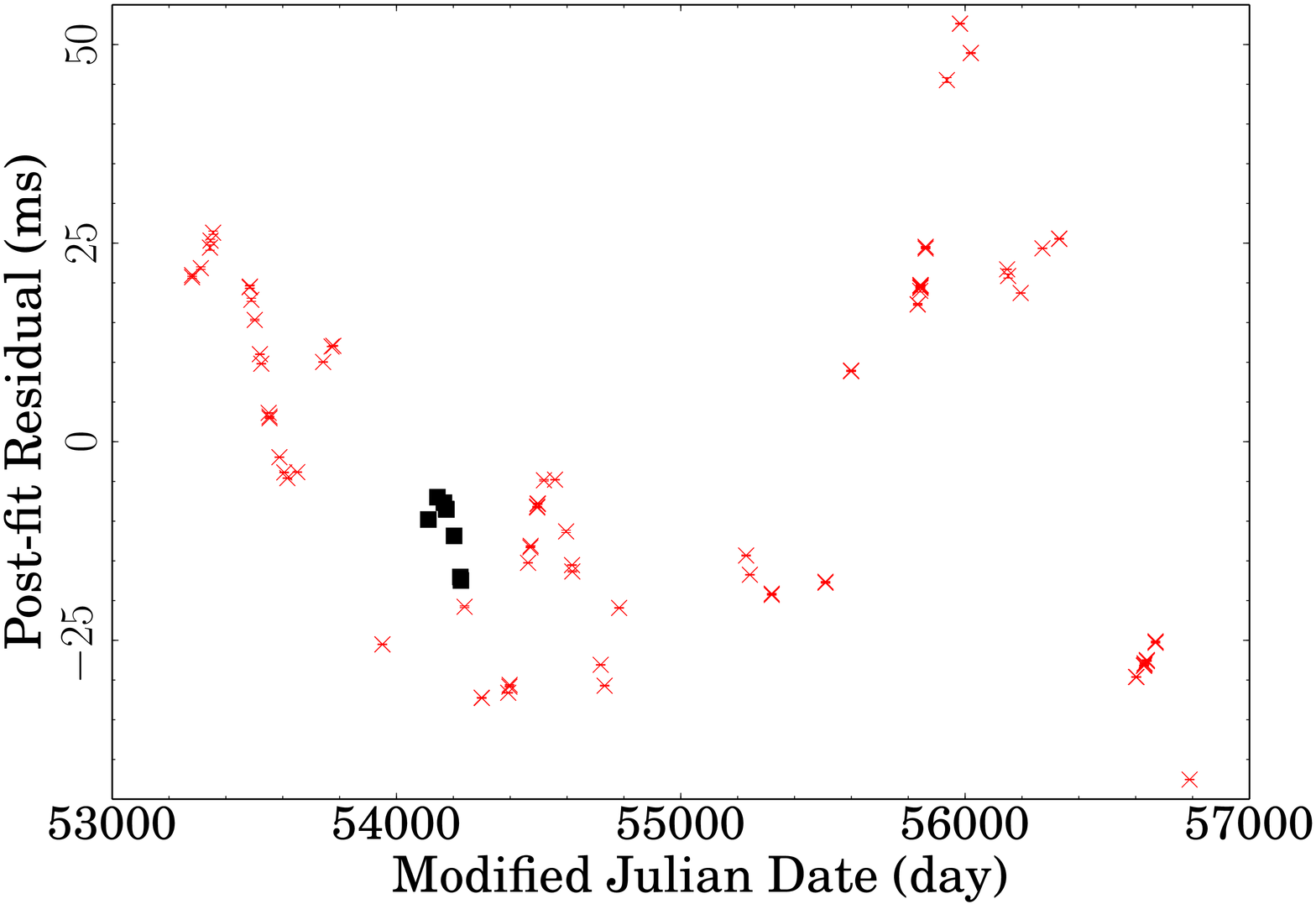}
    \includegraphics[trim= 3mm 1mm -2mm 1mm,clip,angle=270,totalheight=10cm,height=4.5cm,width=8cm]{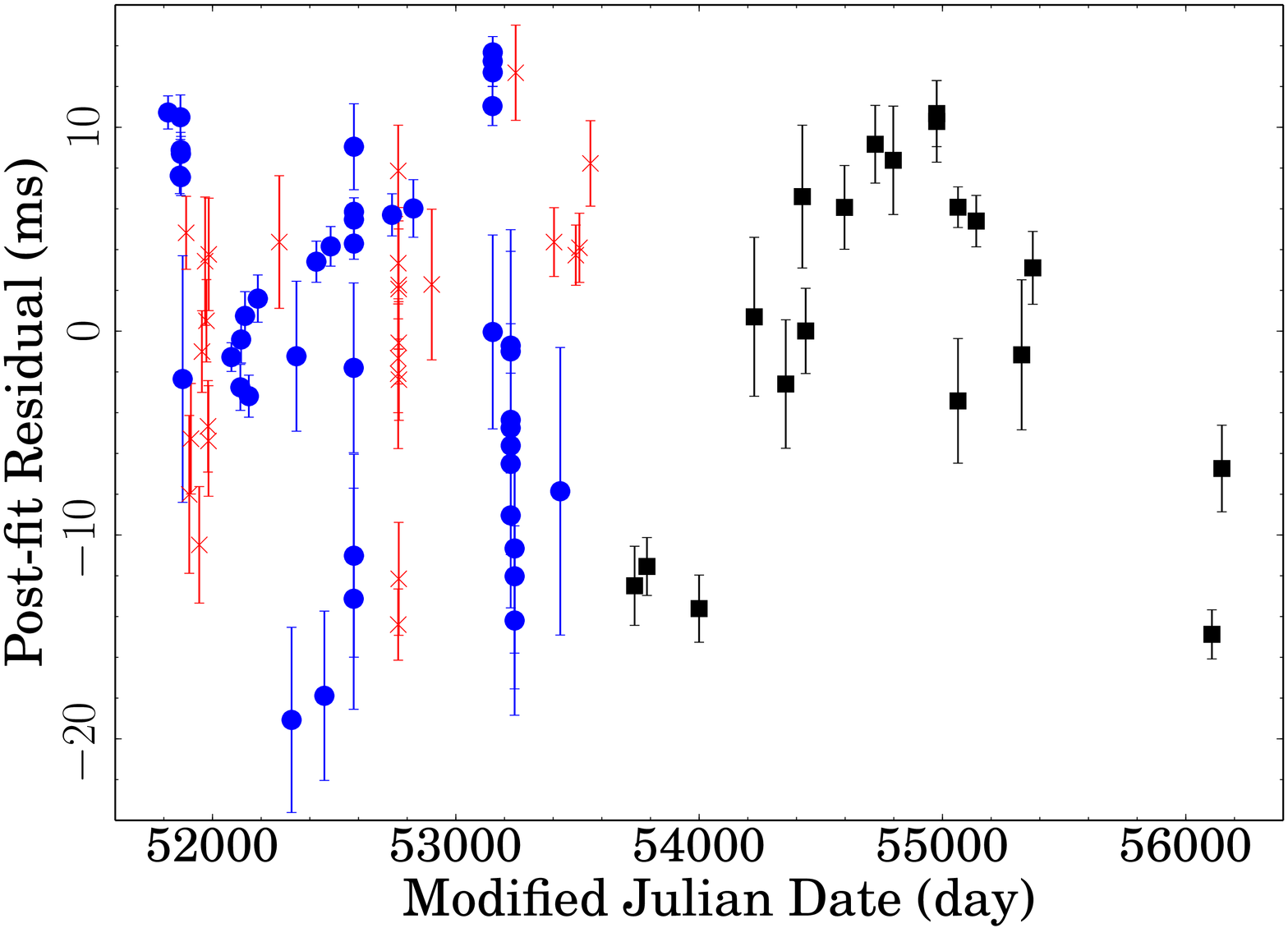}
\vspace{-1mm}
\caption{\emph{Top panel}: best-fit timing residuals for PSR~J1634$-$5107,
  from an 11.1-yr span of observations, with crosses and squares denoting TOAs
  obtained at $\sim$1374~MHz and 1518~MHz, using the Multibeam and H-OH
  receivers, respectively. \emph{Centre panel:} best-fit timing residuals for
  PSR~J1717$-$4054, from 9.6~yr of data, with crosses and squares denoting
  TOAs obtained at 1374~MHz and 1518~MHz, using the Multibeam and H-OH
  receivers, respectively. \emph{Bottom panel:} best-fit timing residuals for
  PSR~J1853$+$0505, from an 11.9-yr long data set, with circles, crosses and
  squares denoting TOAs obtained with the PAFB (1374~MHz and 1518~MHz), LAFB
  (1396~MHz) and LDFB (1402~MHz and 1520~MHz) backends, respectively.}
 \label{res}
\end{figure}

Due to the low number of 10/50cm ON observations, we only considered
observations obtained at L-band for this analysis. Here, we formed analytic
templates from the highest-SNR PAFB (all sources), PDFB (PSRs~J1634$-$5107 and
J1717$-$4054), LDFB (PSR~J1853$+$0505) and LAFB (PSR~J1853$+$0505) profiles
using {\small\textsc{paas}}. Pulse time-of-arrivals (TOAs;
e.g. \citealt{mt77}) were generated through cross-correlation between the
templates and corresponding observations for each source using
{\small\textsc{pat}}. Note that instrumental delays between receivers and/or
backends were also fitted in {\small\textsc{tempo2}}. This process also
ensures that delays caused by inaccurate alignment of the different templates
used for the same source were accounted for. The results of this analysis are
presented in Table~\ref{tab:pars} and Fig.~\ref{res}.

\begin{table*}
\caption{Fitted and derived properties of PSRs~J1634$-$5107, J1717$-$4054 and
  J1853$+$0505 from their best-fit timing solutions. The distance of each
  source is taken from the {\small\textsc{NE2001}} model, and is assumed to
  have a $20\,\%$ uncertainty. Errors are displayed in the parentheses, for
  parameters which were fitted, and are in units of the least-significant
  digit.}  \centering
\begin{tabular}{l l l l l l l l l l l}
  \hline 
  PSR & RA (J2000) & Dec. (J2000) & $\nu$ & $\dot{\nu}$ & DM & d &
  Epoch & $N_{\mathrm{TOA}}$ & $T_{\mathrm{s}}$ & RMS \\
   & (\,h\,:\,m\,:\,s\,) & (\,$^{\circ}$\,:\,$'$\,:\,$''$\,) & (s$^{-1}$) &
  ($10^{-15}$~s$^{-2}$) & (pc~cm$^{-3}$) & (kpc) & (MJD) & & (yr) & ($\mu$s)\\\hline
  J1634$-$5107 & 16:34:04.99(8) & $-$51:07:45.6(9) & 1.97100259922(5) &
  $-$6.1167(5) & 373(2)$^a$ & 6.1 & 54420 & 50 & 11.1 & 4537\\
  J1717$-$4054 & 17:17:51.8(1)  & $-$41:03:20(5)   & 1.12648304468(3) &
  $-$4.6737(7) &  306.9(1)$^b$ & 4.7 & 55035 & 108 & 9.6 & 22011\\
  J1853$+$0505 & 18:53:04.32(4) & $+$05:05:29(1)   & 1.10480469655(3) &
  $-$1.5631(2) & 279(3)$^c$ & 6.6 & 53982 & 89 & 11.9 & 7857\\
  \hline
\end{tabular}
\begin{flushleft}
$^a$Lorimer et al. (2006), $^b$Kerr et al. (2014), $^c$Hobbs et al. (2004).
\end{flushleft}
\label{tab:pars}
\end{table*}

We find that the timing solutions for PSRs~J1634$-$5107 and J1853$+$0505 are
consistent with previous findings \citep{hfs+04,lfl+06}. In contrast to
PSR~J1853$+$0505, we note that the residuals of PSR~J1634$-$5107 are not
white, indicating that $\dot{\nu}$ variations may exist in this
source. However, our observing cadence is unsufficient here to infer anything
conclusive about spin-down variations in the object. We also find substantial
timing noise in PSR~J1717$-$4054 ($\mathrm{RMS}\sim22$~ms;
c.f. \citealt{hlk+04}), which is comparable to that observed in a large number
of sources \citep{hlk10}.

\begin{figure}
  \centering
    \includegraphics[trim= 5mm 2mm 0mm 2mm,clip,angle=270,totalheight=10cm,height=5cm,width=8cm]{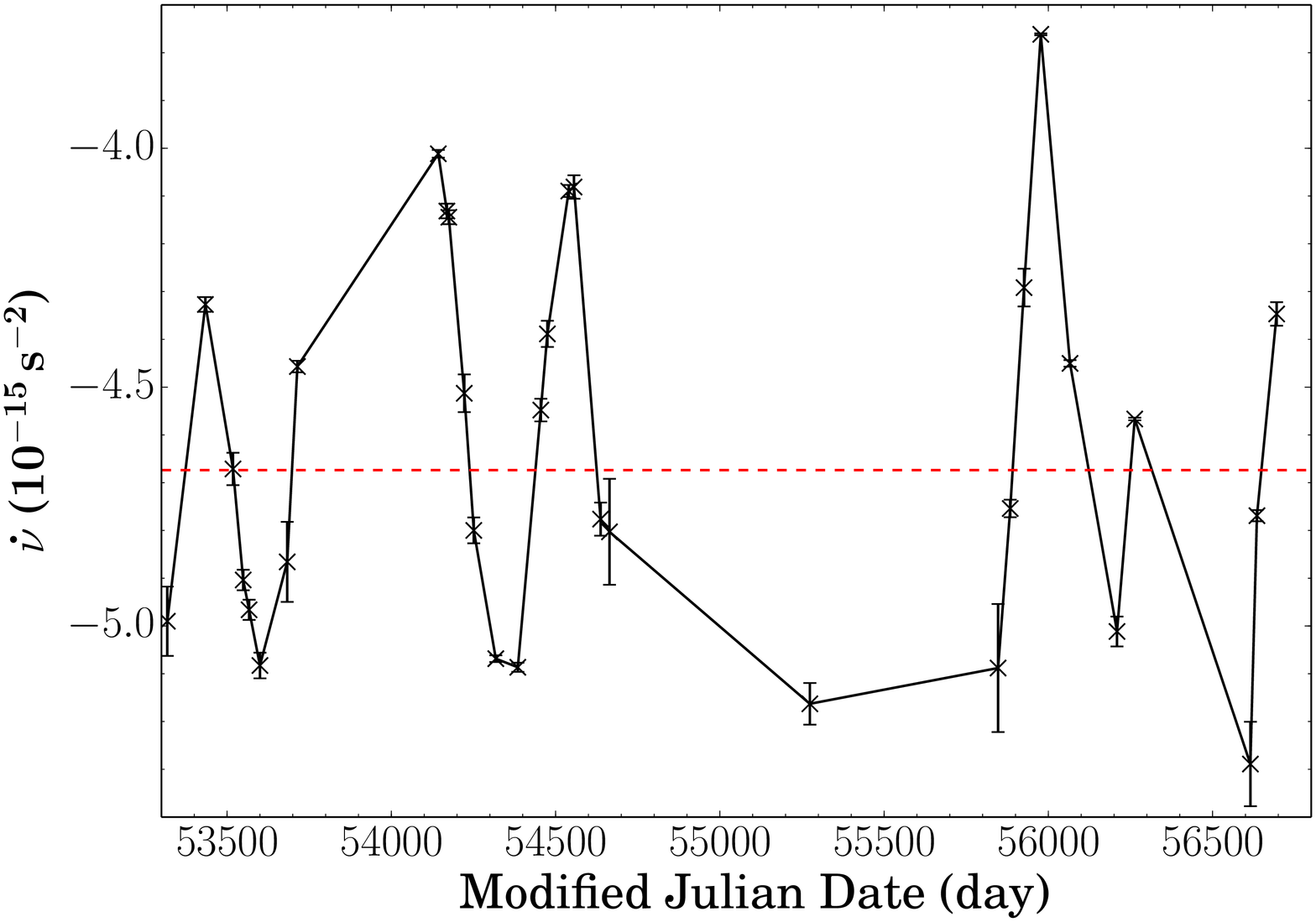}
\vspace{-1mm}
\caption{Variation in $\dot\nu$ against time for PSR~J1717$-$4054, using fit
  windows of 200~d and stride steps of 50~d. The measured
  $\langle\dot\nu\rangle$ from the overall timing solution is also overlaid.}
 \label{nudotevol}
\end{figure}

The timing noise in PSR~J1717$-$4054 can be interpreted as systematic
$\dot\nu$ variations. To quantify this, we performed stride fits across its
timing data using {\small\textsc{tempo2}}, following the method outlined in
\cite{lhk+10}. Here, we used data windows of 200~d in length, and stride steps
of 50~d, to provide the necessary timing accuracy while not compromising too
heavily on time resolution. From this analysis, we find that the source
exhibits a peak-to-peak spin-down variation
$\Delta\nu_{\mathrm{pk}}=1.53\pm0.03\times10^{-15}$~s$^{-2}$ and a fractional
spin-down variation
$|\Delta\nu_{\mathrm{pk}}|/\langle\dot\nu\rangle=33\pm1\,\%$ over the course
of hundreds of days (see Fig.~\ref{nudotevol}).

Following the method outlined in \cite{ysw+12}, we performed WWZ analysis on
the $\dot\nu$ data to determine if the variations were periodic. However, we
find that the cadence of our data is too poor to allow any periodic trend to
be accurately determined. We were also unable to detect any significant pulse
shape variation over time due to the narrow pulse longitude range over which
emission is detected. This precludes any direct correlation between
$\dot{\nu}$ variation and pulse shape variation over time.

\section{Polarimetric Properties}\label{sec:pol}
We also analysed polarimetric Multibeam observations of PSRs~J1634$-$5107 and
J1717$-$4054. As no single pulse observations were available in these data, we
present only the average pulse polarisation properties. The polarisation
calibration was carried out following the scheme outlined in \cite{wj08}.

\subsection{PSR~J1634$-$5107}\label{sec:j1634pol}
We used 12 Multibeam observations to probe the polarimetric properties of
PSR~J1634$-$5107. These were aligned, using the timing solution presented in
Table~\ref{tab:pars}, and then averaged to produce an integrated profile from
the 42~min of data using the {\small\textsc{PSRCHIVE}} software suite. We
determined the rotation measure (RM; see, e.g., \citealt{lk05}) of the source,
$\mathrm{RM}=-151\pm8$~rad~m$^{-2}$, for the first time using the
{\small\textsc{rmfit}} package \citep{njkk08}.

\begin{figure}
\centering
  \includegraphics[trim= 2mm 2mm 1mm 5mm, clip, height=8.15cm,width=5.5cm,angle=270]{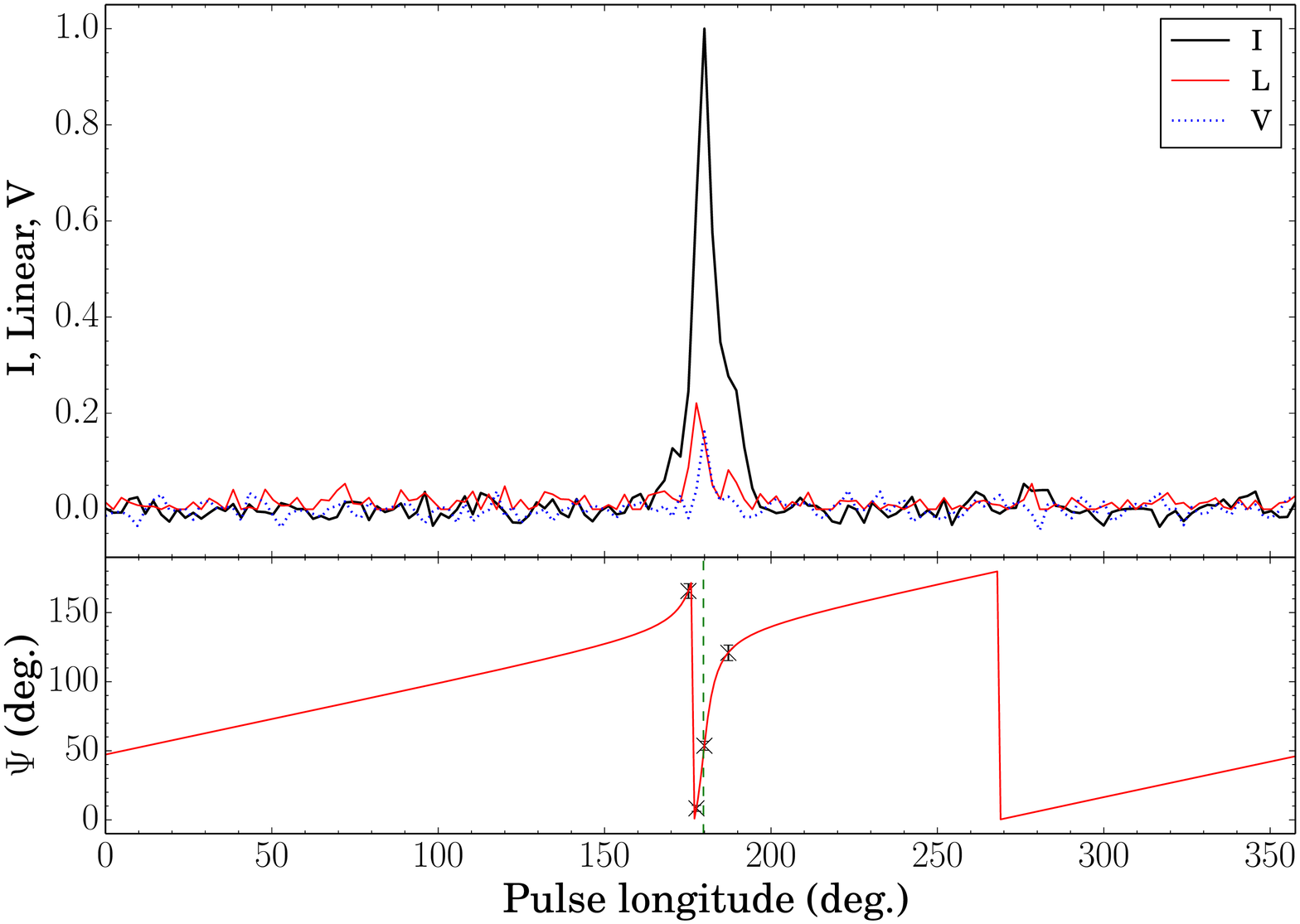} 
\vspace{-1mm}
\caption{The best RVM fit to the integrated 42~min observation of
  PSR~J1634$-$5107. \emph{Top panel:} the average emission properties of the
  integrated observation, showing the total intensity profile, as well as the
  linear and circular polarisation profiles. \emph{Bottom panel:} integrated
  PA data (crosses), with the best-fitting RVM (solid line) and location of
  the PA swing inflection point ($\phi_0$; dotted line) overlaid.}
\label{rvmfit}
\end{figure}

After correcting for RM, we analysed the intrinsic Stokes parameters
($I$,$Q$,$U$,$V$) and degree of linear ($L/I=\sqrt{Q^2+U^2}/I$) and circular
($V/I$) polarisation for the source (see top panel of
Fig.~\ref{rvmfit}). Using a $3\sigma$ limit for the polarisation data, we
found that the pulsar exhibits high linear ($\langle L/I\rangle=19\pm2\,\%$)
and modest circular ($\langle |V|/I\rangle=10\pm1\,\%$) polarisation, when
averaged over bins within the $10\,\%$ intensity pulse width ($W_{10}$)
range. We also computed the maxima of the polarisation profiles;
i.e. $(L/I)_{\mathrm{max}}=40\pm10\,\%$ and
$(|V|/I)_{\mathrm{max}}=30\pm20\,\%$. Here, we quote uncertainties on the
degrees of linear and circular polarisation in terms of their respective
quadrature errors; e.g. $\Delta L = \sigma_{\mathrm{L}}
\sqrt{N_{\mathrm{on}}}$, where $\sigma_{\mathrm{L}}$ is the RMS of $L$ in the
off-pulse region and $N_{\mathrm{on}}$ is the number of on-pulse bins
considered).

Further to the above, we sought to determine the magnetic inclination angle of
the source ($\alpha$) and the impact parameter of the LOS ($\beta$). For this
analysis, we fitted the polarisation position angles (PAs) of the integrated
emission with the rotating vector model (RVM; \citealt{rc69a}), adopting the
$\chi^2$ minimisation technique presented by \cite{rwj15} to obtain fit
constraints with $3\sigma$ PA values.

We find that the best-fit to our data is consistent with the RVM (see
Fig.~\ref{rvmfit}). However, we cannot place rigorous constraints on the
$\alpha$ and $\beta$ parameters due to the low number of significant PA data
points available. Instead, if we assume $\alpha=\ang{90}$, we can estimate the
maximum $\beta$ value ($\beta_{90}$) from the maximum gradient
$(d\Psi$/$d\phi)_{\mathrm{max}}$ of the RVM fit (\citealt{kom70}; see
below). In the general case, $\beta$ is less than this value:
\begin{equation}
\beta_{90} = \mathrm{sin}^{-1}\left[ \mathrm{sin}\,(\ang{90})\,\left( \frac{d\Psi}{d\phi} \right)^{-1}_{\mathrm{max}} \right]\sim\ang{3}\,.
\end{equation}

\subsection{PSR~J1717$-$4054}\label{sec:j1717pol}
\begin{figure}
  \centering
  \includegraphics[trim= 2mm 2mm 1mm 5mm, clip, height=8.15cm,width=5.5cm,angle=270]{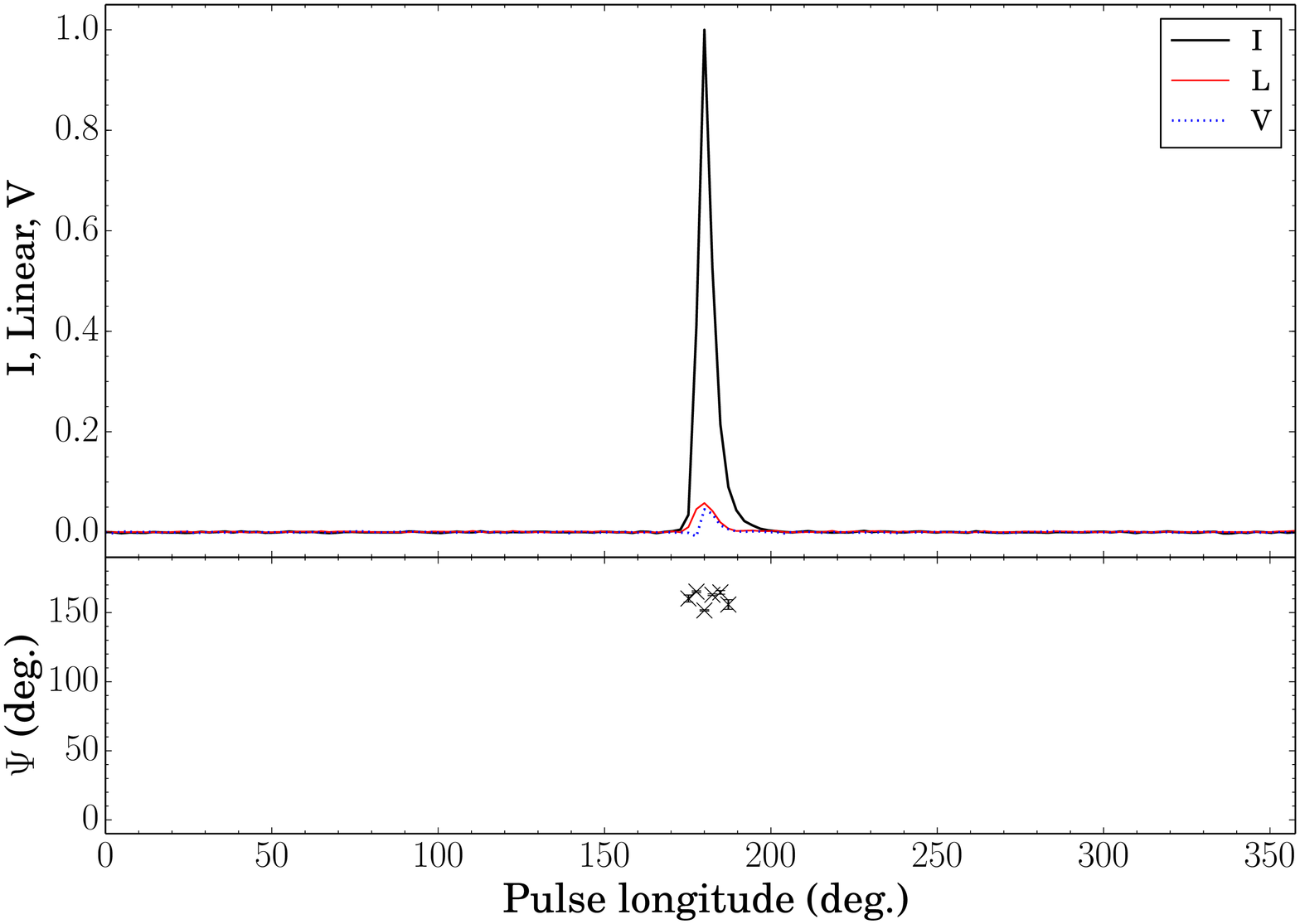} 
\vspace{-1mm}
\caption{The average emission properties of the 146~min observation of
  PSR~J1717$-$4054. \emph{Top panel:} the total intensity (black line), linear
  (red line) and circular (blue line) polarisation profiles. \emph{Bottom
    panel:} integrated PA data for the average profile.}
 \label{1717pol}
\end{figure}

We used 35 Multibeam observations to characterise the polarimetric properties
of PSR~J1717$-$4054. The 146~min of data were aligned and folded, using the
same method as in $\S$~\ref{sec:j1634pol}, to produce an integrated profile
for this analysis. We again used {\small\textsc{rmfit}} to measure an
$\mathrm{RM}=-809\pm3$~rad~m$^{-2}$ for the source. This best-fit RM value,
which represents a refinement of the result of \cite{khs+14}, was then used to
correct the average profile, resulting in the profile shown in the top panel
of Fig.~\ref{1717pol}.

We find that the source exhibits modest linear ($\langle
L/I\rangle=8.1\pm0.1\,\%$) and circular polarisation ($\langle
|V|/I\rangle=4.9\pm0.1\,\%$), when averaged over the $W_{10}$ range (with
$(L/I)_{\mathrm{max}}=30\pm4\,\%$ and
$(V/I)_{\mathrm{max}}=6.9\pm0.2\,\%$). We also attempted to fit the PA data
for the source, using the same method as in $\S$~\ref{sec:j1634pol}, but could
not obtain any constraints on the values of $\alpha$ or $\beta$ from the
resultant reduced $\chi^2$ surface. This is due to the clustering of the PA
data points which are not highly constrained by the RVM (see bottom panel
Fig.~\ref{1717pol}).

\section{Discussion and Conclusions}\label{sec:conc}
\subsection{Emission Variability}\label{sec:varconc}
Thorough analysis of our extensive data set has shown that PSRs~J1634$-$5107,
J1717$-$4054 and J1853$+$0505 exhibit long term nulls and emission phases
(i.e. minutes to many hours), as well as substantial NFs
(i.e. $\sim67\,\%-90\,\%$). Remarkably, PSR~J1853$+$0505 was also shown to
exhibit a weak emission state, in addition to its strong and null states (see
$\S$~\ref{sec:deepnull}).

Short timescale nulling ($\sim1-40~P$) was discovered in PSR~J1717$-$4054,
during its ON phases, which acts to modulate the NF and average detection rate
of the source over sub-integration timescales. This behaviour was also
discovered in an independent analysis performed by \cite{khs+14}, who analysed
the same archival data and two dedicated 9.5-h single-pulse observations. The
latter observing runs constrained a number of mode transitions and thus
allowed tentative limits to be placed on the average ON and OFF emission
timescales of $t_\mathrm{{on}}=1200\pm700$~s and
$t_\mathrm{{on}}=7000\pm5000$~s, respectively. These average emission
durations are found to be consistent with the results obtained in this work,
and indicate that the source exhibits a wide range of emission
timescales. Thus accounting for varied NF reports in the literature.

Overall, the above results place PSRs~J1634$-$5107, J1717$-$4054 and
J1853$+$0505 at the extreme end of the `nulling continuum', where there is a
current deficit of known objects (see, e.g., \citealt{kkl+11,bbj+11}). For
such sources, it is clear that long observations ($\gtrsim1$~h) are required
to best characterise their emission properties, as demonstrated in
\cite{khs+14}.

\vspace{-2mm}
\subsection{Deep Nulls or Weaker Emission States?}\label{sec:deepnull}
PSR~J1853$+$0505 is shown to exhibit three emission states: a weak, strong and
null state. During the weak mode, emission is barely detected above the noise
level in our observations. This behaviour is remarkably similar to that
observed in PSR~J1107$-$5907, where emission at the lowest end of its
weak-mode pulse-energy distribution is easily confused with nulls
\citep{yws+14}. Similar to PSR~J1107$-$5907, integration of $\gtrsim10^3$
pulses is required to detect the particularly weak emission of
PSR~J1853$+$0505, leading to an average
$S_{\mathrm{off}}:S_{\mathrm{weak}}\lesssim 0.11\pm0.03$ at 1374~MHz. These
similarities indicate a close connection between the two objects which, in
turn, advocates further, more detailed analysis of the different emission
modes of PSR~J1853$+$0505.

For the other sources in our study, PSRs~J1634$-$5107 and J1717$-$4054, only
two discrete modes were observed in each object. From this work, we place null
confusion limits of $\lesssim(7\pm2)\times10^{-3}$ and
$\lesssim(1.9\pm0.6)\times10^{-3}$ on these pulsars, respectively. Considering
the fact that PSR~J1634$-$5107 is approximately 3 times fainter than
PSR~J1853$+$0505, it is unsurprising that observations of the former did not
result in the discovery of a weaker emission state. As such, PSR~J1634$-$5107
could exhibit a particularly weak emission state which is not probed by our
observations, or it could actually undergo deep nulls. In the case of
PSR~J1717$-$4054, we suspect that the source exhibits the latter. This is
supported by independent analysis performed by \cite{khs+14}, which resulted
in a more stringent limit on its null confusion limit
(i.e. $\lesssim10^{-3}$). Thus, our observations clearly do not probe a
potential lower intensity emission state in this object.

Since particularly weak emission has been reported in this work, and in a
number of other pulsars
(e.g. $S_{\mathrm{weak}}:S_{\mathrm{strong}}\lesssim2\times10^{-3}$ for
J1107$-$1107; \citealt{yws+14}), it is possible that pulse nulling may only
represent an instrumental sensitivity bias on certain members of the pulsar
population. Overall, these findings provide additional motivation to perform
more regular and longer observations of nulling objects, which will likely
only be possible with next generation telescopes such as MeerKAT and the
SKA. Such steps would ultimately help to construct a census of nulls and mode
transitions in the pulsar population, which is required to help form a fully
consistent pulsar emission model.

\subsection{Timing Properties}\label{sec:timconc}
We have confirmed and updated the timing solutions of PSRs~J1634$-$5107,
J1717$-$4054 and J1853$+$0505, using the longer data spans available in this
work. Both PSRs~J1634$-$5107 and J1853$+$0505 are found to exhibit weak timing
noise, in spite of their extreme variability. Whereas, PSR~J1717$-$4054 is
shown to exhibit modest timing noise, similar to that observed in a number of
pulsars which switch between magnetospheric states \citep{hlk10,lhk+10}.

Upon further investigation of the timing behaviour of PSR~J1717$-$4054, we
found that its timing noise can be accounted for by substantial variation in
$\dot\nu$ over hundreds of days (c.f. \citealt{lhk+10}). This variation is too
slow to be associated with transitions between ON and OFF phases in its
emission. Therefore, it is likely that the source also exhibits emission
variability over similarly long timescales. While we were unable to confirm
this hypothesis through analysis of our data, we predict that more frequent
detections of this source with high-time resolution data will lead to a direct
association with $\dot{\nu}$ variation. Overall, it is clear that
PSR~J1717$-$4054 is a multi-state switching object, which can be used to probe
both short- and long-timescale variations in pulsar magnetospheres.

\subsection{Polarimetric Findings}\label{sec:polconc}
We have presented the first measurement of the RM for PSR~J1634$-$5107, which
has enabled us to fit its $\alpha$ and $\beta$ parameters. While we were
unable to place constraints on $\alpha$, we were able to infer
$\beta_{90}\sim\ang{3}$. Under the assumption that radio beams are comprised
of core and conal emission components, we can determine $\alpha$ from the
observed pulse width of the core ($W^{\mathrm{core}}_{50}$; \citealt{ran90}):
\begin{equation}
\alpha = \mathrm{sin}^{-1}
\left(\frac{\ang{2.45}}{W^{\mathrm{core}}_{50}\,P^{\,0.5}}\right)\,.
\end{equation}
The emission profile of J1634$-$5107, can be described by a strong core
component, flanked by two conal emission components (see
Fig.~\ref{rvmfit}). We thus estimate $W^{\mathrm{core}}_{50}=\ang{6.9}$ and
$\alpha\sim\ang{30}$. However, we stress that longer polarimetric observations
of the source in its ON state, with higher time resolution, are required to
better constrain its emission geometry and verify our results.

We also estimate an $\mathrm{RM}=-809\pm3$~rad~m$^{-2}$ for
PSR~J1717$-$4054. This value is consistent with the result of \cite{khs+14}
and is found to be one of the highest values measured for a pulsar not located
in a globular cluster (see, e.g., \citealt{njkk08} and references
therein). After correcting the source's integrated profile for RM, we also
tried to fit the RVM to its PA data. However, we were unable to obtain any
constraints on $\alpha$ or $\beta$ due to the clustering of PA data points
across the narrow pulse profile, which is again consistent with the result of
\cite{khs+14}. Nevertheless, further investigation into the polarimetric
properties of the source would be interesting to shed light on its
short-timescale variability.

\section{Acknowledgements}
We would like to thank M.~Serylak and the anonymous referee for useful
comments which have improved this manuscript. NJY also acknowledges financial
support from the South African SKA (SKA SA) project.

\bibliographystyle{mn2e}
\bibliography{ver10_arXiv}
\end{document}